\documentclass [12pt,preprint]{aastex} 
\usepackage{epsfig} %
\begin{document} 
\voffset-0.5cm 
\newcommand{\gsim}{\hbox{\rlap{$^>$}$_\sim$}} 
\newcommand{\lsim}{\hbox{\rlap{$^<$}$_\sim$}}

\title{The Anticipated Supernova Associated with GRB090618}

\author{Shlomo Dado\altaffilmark{1} and Arnon Dar\altaffilmark{2}}

\altaffiltext{1}{dado@phep3.technion.ac.il\\
Physics Department, Technion, Haifa 32000,
Israel}
\altaffiltext{2}{arnon@physics.technion.ac.il\\
Physics Department, Technion, Haifa 32000}

\begin{abstract} 

We use the cannonball model of gamma ray bursts (GRBs) and public data 
from the first day of observations of GRB 090618 to predict its X-ray and 
optical lightcurves until very late times, and, in particular, the 
emergence of a photometric and spectroscopic signature of an SN akin to 
SN1998bw in its optical afterglow with an anticipated peak brightness of 
magnitude $\!\sim\!23.2$ in the R band around July 10, 2009, if 
extinction in the host galaxy can be neglected.

\end{abstract}

\section{Introduction}

The relatively nearby bright gamma ray burst (GRB) 090618
that was discovered by the Swift Broad Alert Telescope (BAT)
on June 16, 2009 
at 08:28:29 UT (Schady et al.~2009a)  
provides another good opportunity to investigate the 
association of long duration GRBs with supernova explosions
and to test theoretical models of long GRBs
(e.g. Malesani 2009, Dado and Dar 2009). 
Two such models have been used extensively to analyze GRBs 
and their afterglows (AGs), the fireball (FB) model (for recent reviews 
see, e.g., M\'{e}sz\'{a}ros~2006; Zhang~2007) and the cannonball (CB) 
model (e.g. Dado \& Dar \& De R\'ujula 2002, hereafter DDD002,
Dar \& De R\'ujula~2004, hereafter DD2004; Dado \& Dar \& De 
R\'ujula 2009a,b, hereafter DDD2009a,b, and references therein). 
The two models are quite different and at most only one can 
correctly describe GRBs. Until 
recently, the FB model has been widely accepted as that one. However, 
the rich data on GRBs and their afterglow accumulated from space based 
observations, in particular from the Swift and Fermi satellites, from 
early time observations with ground based robotic telescopes and from 
late-time follow-up observations with large telescopes, have challenged 
this prevailing view (DDD2009a and references therein):  Synchrotron 
radiation (SR), the only radiation mechanism in the original FB 
model, cannot explain simultaneously the prompt optical emission 
and the hard X-ray and gamma-ray emission from GRBs which were both well 
measured in 
some bright GRBs such as 990123 and 080319B. The prompt hard X-ray and 
gamma-ray pulses cannot be explained by synchrotron radiation from 
internal shocks generated by collisions between conical shells. Neither 
can SR explain their typical energy, spectrum, spectral evolution, 
pulse-shape, rapid spectral softening during their fast decay phase and 
the established correlations between various observables. Moreover, the 
early time high energy emission is uncorrelated to the prompt keV-MeV 
emission and lags behind it. As for GRB afterglow (AG), contrary to the 
predictions of the FB model, the broad band AG of GRBs is highly chromatic 
at early times, the AGs of the brightest GRBs do not show jet breaks, and, 
in canonical AGs where breaks are present, they are usually chromatic and 
do not satisfy the closure relations expected from FB model jet breaks.

The situation concerning the CB model is different. The predictions of the 
model which were derived in fair approximations from its underlying 
assumptions were shown to describe correctly the main observed properties 
of GRBs and to reproduce successfully the diverse broadband lightcurves of 
large representative sets of both long GRBs and XRFs (e.g., DDD2009a and 
references therein) and short hard bursts (DDD2009b), in particular of 
relatively nearby bright GRBs with precise and well sampled lightcurves. 
Here we repeat the exercise of predicting from early time observations 
the entire late time behaviour of the afterglow of the recent, relatively 
nearby, bright long GRB 090618 and the emergence of a photometric and 
spectroscopic signature of an SN akin to SN1998bw in its lightcurve before 
their observations.

\section{GRB090618}

At 08:28:29 UT, the Swift Burst Alert Telescope (BAT) triggered and 
located the bright long duration gamma ray burst (GRB) 090618 (Schady et 
al.~2009a) at redshift $z$=0.54 (Cenko et al.~2009b).  About 90\% of the 
GRB energy measured by BAT was emitted within T90=113s (Baumgartner et 
al.~2009). The Swift X-ray telescope (XRT) began follow up observations of 
its X-ray lightcurve (see Fig.~1) 124 s after the BAT trigger and its UVO 
telescope detected its optical AG 129s after trigger (Schady et 
al.~2009b ). The burst was detected also by AGILE (Longo et al.~2009), 
Fermi GBM (McBreen et al.~2009), Suzaku WAM (Kono et al.~2009), KONUS-WIND 
and KONUS-RF (Golenetskii et al.~2009). The burst light curve showed a 
smooth multipeak structure with 4 prominent peaks (one followed by 
three much 
brighter overlapping peaks)  with a total duration of ~160 s. 
Significant spectral evolution was observed during the burst. The 
spectrum at the maximum count rate, measured from T+62.720 to T+64.0 s, 
was well fitted (Golenetskii et al.~2009) in the 20 keV - 2 MeV range by 
the Band function (Band et al.~1993) with a low-energy photon index -0.99 
(-0.06, +0.07), a high energy photon index -2.29(-0.5, +0.23), and peak 
energy Ep=440$\!\pm\!$70 keV while the time integrated  spectrum 
had a low-energy photon index -$1.28\!\pm\!.02\,,$ a high energy photon 
index -2.66(-0.2, +0.14), and a peak energy Ep=186$\pm\!$8 keV. The 
isotropic equivalent energy in the 8-1000 keV band was $E_{iso}\!=\!2.0\times 
10^{53}$ erg (standard cosmology).

The bright optical afterglow of GRB090618 was first detected by the 
ROTSE III robotic telescope 23.9 s after the BAT trigger (Rujopakarn et 
al.~2009) and by the Palomar 60-inch telescope (Cenko et al.~2009a), the 
Katzman Automatic Imaging Telescope (Perley et al.~2009, Li et al.~2009) 
and the UVO Telescope aboard Swift (Schady et al.~2009a) within 2 minutes 
after trigger. 
Absorption features which were detected in its bright optical AG with the 
3m Shajn telescope at Lick observatory yielded a redshift of $z$=0.54. Its 
optical AG was followed up by many telescope and reported shortly after in 
GCN circulars (see Table I). Its R band light curve  reported in 
these GCN circulars  before June 27, 2009 is shown in Fig.~2~.

\section{The CB model}
 
In the cannonball (CB) model (DDD2002, DD2004, DDD2009a, and references 
therein) long-duration GRBs and their AGs are produced by bipolar jets of 
highly relativistic CBs of ordinary matter which are ejected (Shaviv \& 
Dar~1995, Dar \& Plaga~1999) in core-collapse supernova (SN) explosions 
akin to SN 1998bw (Galama et al.~1998). Their prompt MeV gamma-rays and 
hard X-rays are produced by the thermal electrons in the CBs' plasma by 
inverse Compton scattering (ICS) of glory photons - photons 
emitted/scattered into a cavity created by the wind/ejecta blown from the 
progenitor star prior to the SN. When the CBs cross the wind/ejecta and 
coast through the interstellar medium (ISM) behind it, the electrons of 
the ionized gas in front of them that are swept in and Fermi accelerated 
by the CBs' turbulent magnetic fields emit synchrotron radiation (SR) 
which dominates the `prompt' optical emission and the broad band afterglow 
emission. ICS of the SR radiation by these electrons and the decay of 
$\pi^0$'s produced in collision between the swept-in wind and ISM protons 
and the ambient CB protons produce the `prompt' high energy emission 
simultaneously with the optical emission.  Within the CB model, the above 
radiation mechanisms with the burst environment suffice to provide a 
sufficiently accurate description of the observed radiations from GRBs at 
all times and all detected wavelengths.

\subsection {The optical lightcurve}

In the CB model, the observed optical light has three origins: the 
ejected CBs, the SN explosion, and the host galaxy.  
The optical light curve is the sum of 
their energy flux density:
\begin{equation} 
F_{AG}[\nu,t] = F_{CB}[\nu,t] + F_{SN}[\nu,t] + F_{HG}[\nu,t]\,.
\label{sum}
\end{equation}
The contribution of the host galaxy, $F_{HG}$, is usually extracted from 
very late time observations when the CB and SN contributions become 
negligible. In the case of GRB090618, 
a faint object is visible in the SDSS r and i frames at the 
position of the optical afterglow, likely its host galaxy.
Compared to nearby SDSS stars, its r magnitude was estimated
by Malesani (2009) to be $22.7\!\pm\!0.3$.

The energy flux density 
of an SN like SN1998bw with an energy flux density $F_{bw}[\nu,t]$
at redshift $z_{bw}\!=\!0.0085$ (Galama et al.~1998) 
placed at a redshift $z$ is given by 
(e.g., DDD2002),
\begin{equation}
F_{SN}[\nu,t]=k\, {D_L^2(z_{bw})\over D_L^2(z)}\,
{A_{SN}[\nu,z]\over A_{BW}[\nu, z_{bw}]}\,F_{bw}[k\nu, t/k]\, ,
\label{FSN}
\end{equation}
where $k\!=\!(1\!+\!z)/(1\!+\!z_{bw})$,   
$A(\nu,z)$ 
is the attenuation of the observed SN light at frequency $\nu$ arriving 
along the line of sight, and $D_L(z)$ is the luminosity 
distance to redshift z (we use the standard cosmology with 
$\Omega_M\!=\!0.27$,
$\Omega_\Lambda\!=\!0.73$  and $H_0\!=\!71\, {\rm km\, s^{-1}\,  
Mpc^{-1}}$). 

In the CB model (DDD2009a,b and references therein) the afterglow emission 
begins when the CBs encounter the wind/ejecta of the progenitor star. It 
is dominated by synchrotron radiation (SR) from the electrons of the 
ionized 
wind and interstellar-medium in front of the CBs which are swept in and 
Fermi accelerated by the CBs' turbulent magnetic fields which we assume to 
be in approximate energy equipartition with their energy.

The SR, isotropic in the CB's
rest frame, has a characteristic frequency, $\nu_b(t)$,
the typical frequency radiated by the
electrons that enter a CB at time $t$ with a relative Lorentz
factor $\gamma(t)$, and spiral around its equipartition 
magnetic field  (with the incident 
protons). In the observer's frame:
\begin{equation}
\nu_b(t)\simeq  {\nu_0 \over 1+z}\,
{[\gamma(t)]^3\, \delta(t)\over 10^{12}}\,
\left[{n\over 10^{-2}\;\rm cm^3}\right]^{1/2}
{\rm Hz},
\label{nub}
\end{equation}
where $\nu_0\!\simeq\! 3.85\times 10^{16}\, \rm Hz \simeq 160\, eV$
and $\delta(t)$ is the Doppler factor of the CB.
The spectral energy density of the SR
from a single CB at a luminosity distance $D_L$  is given by (DDD2009a):
\begin{equation}
F_{CB}[\nu,t] \simeq {\pi\, R^2\,n\, m_e\, c^3\,
\gamma(t)^2\, \delta(t)^4\, A(\nu,t)\,
\over 4\,\pi\, D_L^2\,\nu_b(t)}\;{p-2\over p-1}\;
\left[{\nu\over\nu_b(t)}\right]^{-1/2}\,
\left[1 + {\nu\over\nu_b(t)}\right]^{-(p-1)/2}\,,
\label{Fnu}
\end{equation}
where $p\!\sim\! 2.2$ is the typical spectral index
of the Fermi accelerated
electrons, and $A(\nu, t)$
is the attenuation of photons of observed frequency $\nu$ along the line
of sight through the CB, the host galaxy (HG), the intergalactic medium
(IGM) and the Milky Way (MW):
\begin{equation}
A(\nu, t) = {\rm
exp[-\tau_\nu(CB)\!-\!\tau_\nu(HG)\!-\!\tau_\nu(IGM)\!-\!\tau_\nu(MW)].}
\label{attenuation}
\end{equation}
The opacity $\tau_\nu\rm (CB)$ at very early times, during the
fast-expansion phase of the CB, may strongly depend on time and frequency.
The opacity of the circumburst medium [$\tau_\nu\rm (HG)$ at early times]
is affected by the GRB and could also be $t$- and $\nu$-dependent.  The
opacities $\tau_\nu\rm (HG)$ and $\tau_\nu\rm (IGM)$ should be functions
of $t$ and $\nu$, for the line of sight to the CBs varies during the AG
observations, due to the hyperluminal motion of CBs.

\subsection{The early-time SR}
At early-time, 
before the CB has swept a mass comparable to its rest mass 
both $\gamma$
and $\delta$ stay put at their initial values
$\gamma_0$ and $\delta_0$.
Then, Eq.~(\ref{Fnu})
yields an early-time SR light curve,
$F_{SR}[\nu,t]\! \propto\! e^{-\tau_{_W}}\, R^2\, 
n^{(1+\beta)/2}\,\nu^{-\beta}.$
Since $r\!\propto\!t$,
a CB ejected into
a windy density profile, $n\!\propto\!1/r^2$,
created by the mass ejection from the
progenitor star prior to its SN explosion, emits SR
with an early-time light curve,
\begin{equation}
F_{SR}[\nu,t] \propto  {e^{-a/t}\,
t^{1-\beta} \over t^2+t_{exp}^2}\, \nu^{-\beta}\!\rightarrow
        \! t^{\!-\!(1\!+\!\beta)}\, \nu^{\!-\!\beta}\, .
 \label{SRP}
\end{equation}
For a CB ejected at time $t_i$, the time  $t$ must be replaced by
$t\!-\!t_i$, the time after ejection.
In the $\gamma$-ray and X-ray bands, the SR emission from a CB is usually
hidden under the prompt keV-MeV ICS emission. But, in many GRBs, the 
asymptotic
exponential decline of the energy flux density of the prompt 
keV-MeV ICS emission
is taken over by the slower power-law decline, 
$F_{SR}[\nu,t]\! \propto\! t^{-\Gamma_X}\, \nu^{-\Gamma_X\!+\!1}$ 
with $\Gamma_X\!=\!\beta_X\!+\! 1\!\approx\! 2.1 $ of the
synchrotron emission in the windy $\!\sim\! 1/r^2$ circumburst density
before the CB reach the constant ISM density and
the AG enters  
a plateau phase  (see examples, e.g., in DDD2009a). Note that 
the `prompt' optical emission that is dominated 
by SR, decays initially like $F_{SR}[\nu,t]\! \propto\! 
t^{\!-\!1.5}\,\nu^{\!-\!0.5}$ since  
the spectral index of the unabsorbed SR emission 
at frequencies well below 
the bend frequency is $\beta_O\!\approx\!\!0.5\,.$
As the wind density decreases with distance, 
the bend frequency may cross the optical band while the CB is still in the 
wind, yielding a steeper decay,  
$F_{SR}[\nu,t]\!\rightarrow \!t^{\!-\!2.1}\,\nu^{\!-\!1.1}\,.$

\subsection{The plateau, the break and the late time decay}

During the coasting phase of a CB in a constant density ISM
the behaviour of its SR lightcurve as given by Eq.~(\ref{Fnu}) is 
dominated by the time dependence 
of its Lorentz factor $\gamma(t)$. 
From energy-momentum  conservation it follows that 
\begin{equation}
\gamma(t) = {\gamma_0\over [\sqrt{(1+\theta^2\,\gamma_0^2)^2 +t/t_0}
          - \theta^2\,\gamma_0^2]^{1/2}}\,,
\label{goft}
\end{equation}
with
\begin{equation}
t_0={(1\!+\!z)\, N_{_{\rm B}}\over 8\,c\, n\,\pi\, R^2\,
\gamma_0^3}\,.
\label{break}
\end{equation}
This deceleration law is for the
case in which the ISM particles re-emitted fast by the
CB are a small fraction of the flux of the intercepted ones.
As can be seen from Eq.~(\ref{goft}), $\gamma$  and $\delta$
change little as long as $t\!\ll\! t_b\!=\![1\!+\gamma_0^2\,
\theta^2]^2\,t_0\, $ which results in the shallow decline/plateau phase of 
the AG. 
In terms of typical CB-model values of $\gamma_0$,
$R$, $N_{_{\rm B}}$ and $n$,
\begin{equation}
t_b= (1300\,{\rm s})\, [1+\gamma_0^2\, \theta^2]^2\,(1+z)
\left[{\gamma_0\over 10^3}\right]^{-3}\,
\left[{n\over 10^{-2}\, {\rm cm}^{-3}}\right]^{-1}
\left[{R\over 10^{14}\,{\rm cm}}\right]^{-2}
\left[{N_{_{\rm B}}\over 10^{50}}\right] \! .
\label{tbreak}
\end{equation}
For $t\!\gg\!t_b$, $\gamma$  and $\delta$ decrease like $t^{-1/4}\,.$
The transition $\gamma(t)\!\sim\! \gamma_0\! \rightarrow\!\gamma\!\sim\!
\gamma_0\,(t/t_0)^{-1/4}$
induces a bend (the so called `jet  break')
in the synchrotron AG from a plateau to an asymptotic power-law
decay,
\begin{equation}
F_{SR}[\nu,t] \propto t^{-p/2-1/2}\,\nu^{-p/2}= t^{-\beta-1/2}\,
\nu^{-\beta}=t^{-\Gamma+1/2}\,\nu^{-\Gamma+1},
\label{Asymptotic}
\end{equation}
with a power-law in time steeper by half a unit
than that in frequency.

\section{The X-ray lightcurve of the afterglow of GRB090618}

The X-ray lightcurve of GRB090618 (Evans et al.~2008)
shows the canonical behaviour predicted by the CB model (e.g. DDD2002; 
DDD2009a) and displayed by many Swift GRBs (e.g., Nousek et al.~2006).
This behaviour is well reproduced by the CB 
model as shown in Fig.~1. The fast decline phase 
with a rapid spectral softening 
is that predicted for ICS of glory light
(e.g. Dado, Dar  and De R\'ujula 2008a; DDD2009a):
$F_{ICS}[\nu,t]\!\rightarrow\! t^{-2}\, e^{-E\, t^2/E_p\, t_p^2}$.
where $t$ and $t_p$ are measured relative to 
the beginning of the last large prompt emission 
episode (we used a best fit value, $t_p\!=\!12.9$). 
The sharp transition around 300 s to a shallow decline/plateau phase 
with a constant hardness ratio is produced 
when the synchrotron afterglow  given by 
Eq.~(\ref{Fnu}) takes over. 
The shape of the SR afterglow of GRB090618 was reproduced 
with three best fitted parameters, 
$\gamma_0\, \theta\!=\!1.10$ and $t_0\!=\!312$s
which yield $t_b\!=\! 1540$s and $p\!=\!2.08$, 
This best fit value of $p$ satisfies well the CB model relation, 
$p\!=\!2\Gamma\!-\!2\!=\!2.02\!\pm\!0.10$, 
where we used the photon spectral index 
$\Gamma$=2.008 (+0.047, -0.046) that was reported in the Swift repository 
(Evans et al.~2008).
The `jet break' takes place 
when the jet of CBs gathers a mass comparable to 
its rest mass (DDD2002; DDD2008b; DDD2009a)
as given by Eqs.~(\ref{break}) and (\ref{tbreak}). 
The post-break power-law 
decay of the AG is well described (DDD2008b) by Eq.~(\ref{Asymptotic}).

\section{The optical AG of GRB 090618 and emergence of an underlying SN?}
   
The late time optical AG can be estimated by extrapolating the post break 
power-law decay to the time of the anticipated emergence of an SN 
signature akin to that of SN1998bw at the burst location.  However, unlike 
the X-ray light curve which was inferred from a continuous follow-up 
measurements with the same telescope (Swift XRT), the optical lightcurve 
of GRB090618 constructed from reported measurements in GCNs with different 
telescopes at different times, locations, atmospheric and seeing 
conditions, calibrations and spatial resolutions. In particular, the 
detection of the SN signature depends on a precise subtraction of the host 
galaxy contribution to the observed lightcurve. In view of all that we 
preferred to best fit the early-time observational data on the optical 
afterglow which were reported in the GCNs listed in Table I and to use the 
CB model with the parameters determined from the X-ray AG and the early 
time observational data (Table I) on the optical AG of GRB090618 to 
predict the late time R-band lightcurve of the optical transient. A host 
galaxy contribution of r=$22.7\!\pm \! 0.3$ (Malesani et al.~2009) was 
subtracted from the last two data points, and the anticipated 
SN1998bw-like contribution at the host location was dimmed by the Galactic 
extinction along the line of sight corresponding to E(B-V)=0.09. The 
results of this exercise are presented in Fig.~2.

Although the very early optical emission does not directly affect the 
late-time behaviour of the AG, we have also fitted the very early 
(`prompt') optical emission in a windy environment as given by 
Eq.~(\ref{SRP}), for completeness and in order to demonstrate the validity 
of the CB model. The fit to the two prompt emission peaks was obtained 
with $\beta_O$=0.50, $t_i$=25.1s, $a$=1.47s, $t_{exp} $=19.4s for the 
first peak and $t_i$=52.4s, $a$=98s, $t_{exp}$=40.3s for the second.  The 
scarcity of data points during the rise part of the peaks allows for other 
equally good fits.

A careful subtraction of the host galaxy contribution from the optical 
lightcurve of GRB090618 measured with large telescopes of good spatial 
resolution should have shown already (on June 29, 2009) spectroscopic and 
photometric evidence for an underlying supernova if its brightness is 
similar to that of SN1998bw. As of today, the R band lightcurve should 
show a plateau with a very small rise towards a shallow maximum around 
July 10, when an SN akin to SN1998bw with only little extinction in the 
host galaxy will reach its peak brightness of magnitude 
$\!\sim\!23.2$ in the R band.  
The detailed data on the broadband afterglow of GRB090618, which is less 
bright than that of GRB030329, can still provide another useful test of 
the long-duration GRB-SN association and of GRB models.

\begin{deluxetable}{llllc}
\vskip -2.cm
\tablewidth{0pt}
\tablecaption{GCNs Reporting Follow-up R-band Optical Observations of 
GRB090618  Before June 25, 2009} 
\tablehead{
\colhead{GCN} & \colhead{Telescope} & 
\colhead{Reported by} &\colhead{Comments}
}
\startdata
9512 & Swift UVOT      &  Schady et al. & Finding exposure    \\
9513 & Palomar 60-inch &  Cenko et al.  &    \\
9514 & KAIT  & Perley et al. &    \\
9515 & ROTSE-III   & Rujopakarn et al. & Earliest Detection    \\
9517 & KAIT  & Li et al. & Follow-up    \\
9518 & Lick 3-m &  Cenko et al. & Redshift Measurement   \\
9519 & OAGH 2.1-m  &  Carraminana et al. & Follow-up   \\
9520 & Faulkes  &  Melandri et al.  & Follow-up    \\
9522 & Mt. Lemmon 1-m & Im. et al. & Follow-up   \\
9526 & SDSS  &  Malesani & Host Detection   \\
9531 & Liverpool  & Cano et al.  & Follow-up   \\
9539 & Shajn  &  Rumyantsev et al. & Follow-up  \\
9542 & SAO RAS 6-m  & Fatkhullin et al.  &  Spectroscopy   \\
9548 & RTT 1.5-m  & Galeev et al.  & Follow-up   \\
9575 & SARA 0.9-m  & Updike et al.  & Follow-up   \\
9576 & Himalayan 2-m  & Anupama et al. & Follow-up   \\
\enddata
\label{t1}
\end{deluxetable}

\newpage
\begin{figure}[]
\centering
\epsfig{file=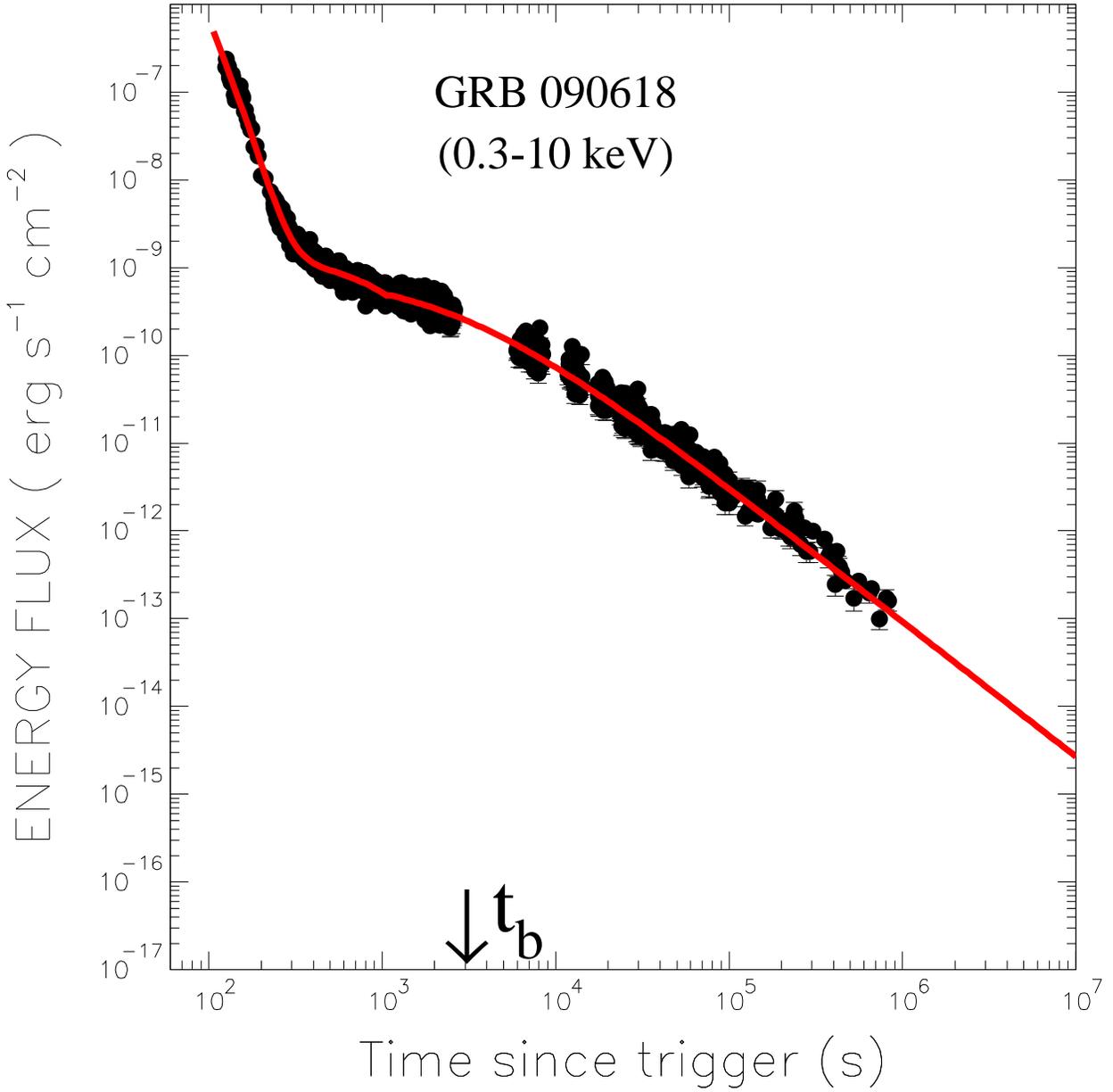,width=18cm}
\caption{Comparison between the Swift XRT lightcurve of
GRB090618 (Evans et al.~2009)
and its CB model description 
(see text).
The time of the deceleration
bend/break $t_b$ is indicated by a down-pointing arrow.}
\label{fig1}
\end{figure}

\newpage
\begin{figure}[]
\centering
\epsfig{file=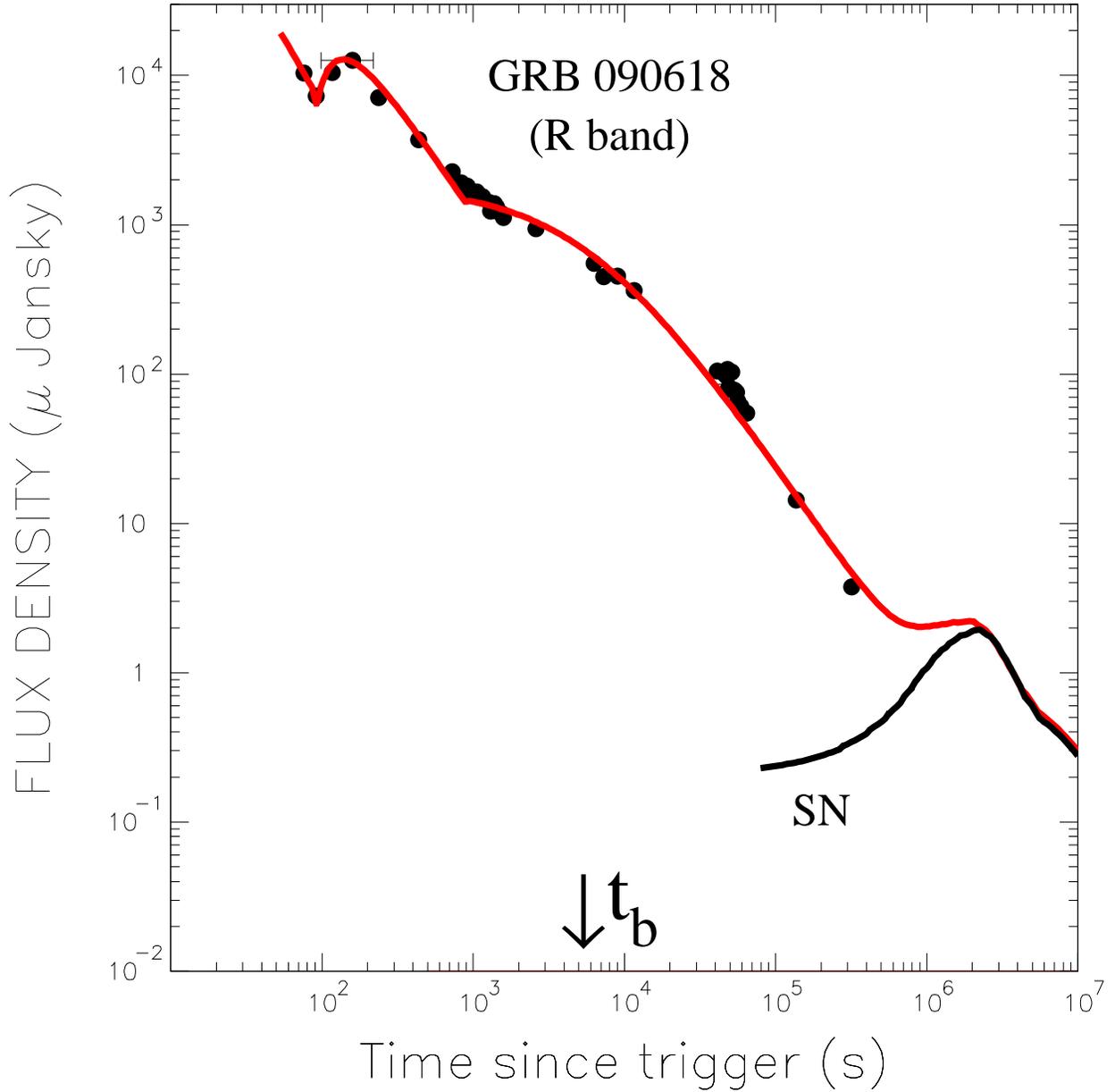,width=18cm}
\caption{Comparison between the R band lightcurve of
GRB090618 extracted from the GCN circulars 
listed in Table I and its CB model description in terms of the 
parameters which were extracted from the CB model fit 
to its early X-ray lightcurve measured by the Swift XRT
(Evans et al.~2009). The plotted SN lightcurve is that of 
SN1998bw at the burst location. Its maximum brightness is expected 
around July 10.}
\label{fig2}
\end{figure}

\end{document}